\documentclass[aps,pra,reprint,groupedaddress,showpacs]{revtex4-1}

\usepackage{graphicx,amssymb,amsmath}
\usepackage{color}
\definecolor{orange}{rgb}{0.7,0.3,0.0}
\definecolor{green}{rgb}{0.1,0.6,0.0}

\begin{document}


\title{Superconducting switch for fast on-chip routing of quantum microwave fields}


\author{M. Pechal}
\email[]{mpechal@phys.ethz.ch}
\author{J.-C. Besse}
\author{M. Mondal}
\author{M. Oppliger}
\author{S. Gasparinetti}
\author{A. Wallraff}
\affiliation{Department of Physics, ETH Zurich, CH-8093 Zurich, Switzerland}


\date{\today}

\begin{abstract}
A switch capable of routing microwave signals at cryogenic temperatures is a desirable component for state-of-the-art experiments in many fields of applied physics, including but not limited to quantum information processing, communication and basic research in engineered quantum systems. Conventional mechanical switches provide low insertion loss but disturb operation of dilution cryostats and the associated experiments by heat dissipation. Switches based on semiconductors or microelectromechanical systems have a lower thermal budget but are not readily integrated with current superconducting circuits. Here we design and test an on-chip switch built by combining tunable transmission-line resonators with microwave beam-splitters. The device is superconducting and as such dissipates a negligible amount of heat. It is compatible with current superconducting circuit fabrication techniques, operates with a bandwidth exceeding $100\,\mathrm{MHz}$, is capable of handling photon fluxes on the order of $10^{5}\,\mu\mathrm{s}^{-1}$, equivalent to powers exceeding $-90\,\mathrm{dBm}$, and can be switched within approximately $6-8\,\mathrm{ns}$. We successfully demonstrate operation of the device in the quantum regime by integrating it on a chip with a single-photon source and using it to route non-classical itinerant microwave fields at the single-photon level.
\end{abstract}


\maketitle








\section{Introduction}

Many fields of research ranging from material physics to quantum information processing make use of microwave measurements at temperatures as low as few tens of millikelvin. To build microwave systems capable of operation at these low temperatures often requires the use of components specifically designed for such conditions. For active devices such as amplifiers, the usual semiconductor-based designs dissipate too much power to be compatible with millikelvin temperatures \cite{Pospieszalski1988}, necessitating alternative approaches such as superconducting devices based on parametric amplification \cite{Yurke2006,Castellanos2007}.

An RF switch is another example of a device commonly used in room-temperature microwave systems, whose integration in cryogenic setups is problematic. Latching mechanical switches have been successfully used for setups inside dilution refrigerators \cite{Ranzani2013} but the heat-load due to the solenoids used for moving the mechanical parts causes significant increase in temperature after each switching event and a rather long time of approximately $15\,\mathrm{min}$ is needed for the cryostat to reach its base temperature again. PIN diode switches which need to be continuously current-biased during operation are even less suitable for low-temperature applications. The quickly developing family of MEMS microwave switches \cite{Rebeiz2001,Schoenlinner2011} can provide very reliable switching with a minimal power dissipation and seems to be a good candidate for cryogenic microwave setups \cite{Gong2009,Attar2014}. Similarly promising devices were recently demonstrated based on field-effect transistors \cite{AlTaie2013,Ward2013,Puddy2015,Hornibrook2015}. However, these types of switches cannot be easily combined with on-chip superconducting circuits without a significant modification of the fabrication process.

On-chip switching devices will be a useful tool for construction of more complex integrated quantum systems and have a potential to enable a range of novel experiments. For example, in combination with measurement-based entanglement generation techniques, such as demonstrated in Ref.~\cite{Narla2016}, they could be used to create distributed multi-qubit entangled states. In experiments of the Hong-Ou-Mandel type \cite{Lang2013}, microwave field states could be routed in a network of switches and beam-splitters to generate a range of different non-classical states using a single device. The ability to route signals on-chip can also provide a convenient way to calibrate a linear detection setup by fabricating a suitable source of a reference signal on the same chip as the device-under-test.

Here we describe and experimentally test a microwave switch design which is integrated on a superconducting chip using fabrication procedures widely employed in the field of superconducting quantum circuits. Devices of a similar type were recently implemented \cite{Naaman2016,Chapman2016b} but are currently restricted to the single-pole, single-throw (SPST) mode of operation. That is, they can either route a signal towards an output port or reflect it back to the input. In contrast, our switch is of the single-pole, double-throw (SPDT) type -- it switches a single input to one of two outputs or vice versa. The device is based on interference effects in a microwave circuit and has no moving parts. The expected internal loss rate in a superconducting circuit of this type \cite{Goppl2008} is at least three orders of magnitude lower than the 3dB bandwidth, which is on the order of $100\,\mathrm{MHz}$ in the device studied here. This bandwidth is smaller than that of mechanical or PIN diode switches but sufficient for many applications encountered in superconducting circuit experiments. The device is controlled by externally applied magnetic flux which, if generated by an on-chip flux line, can be tuned on time-scales of a few nanoseconds \cite{Sandberg2008a}, making the switching very fast as compared to typical coherence times of state-of-the-art superconducting circuits. This will make it useful in applications where routing of signals needs to be controlled by real-time feedback.

\section{Principle of operation}

\begin{figure*}
\begin{center}
\includegraphics[width=17.5cm]{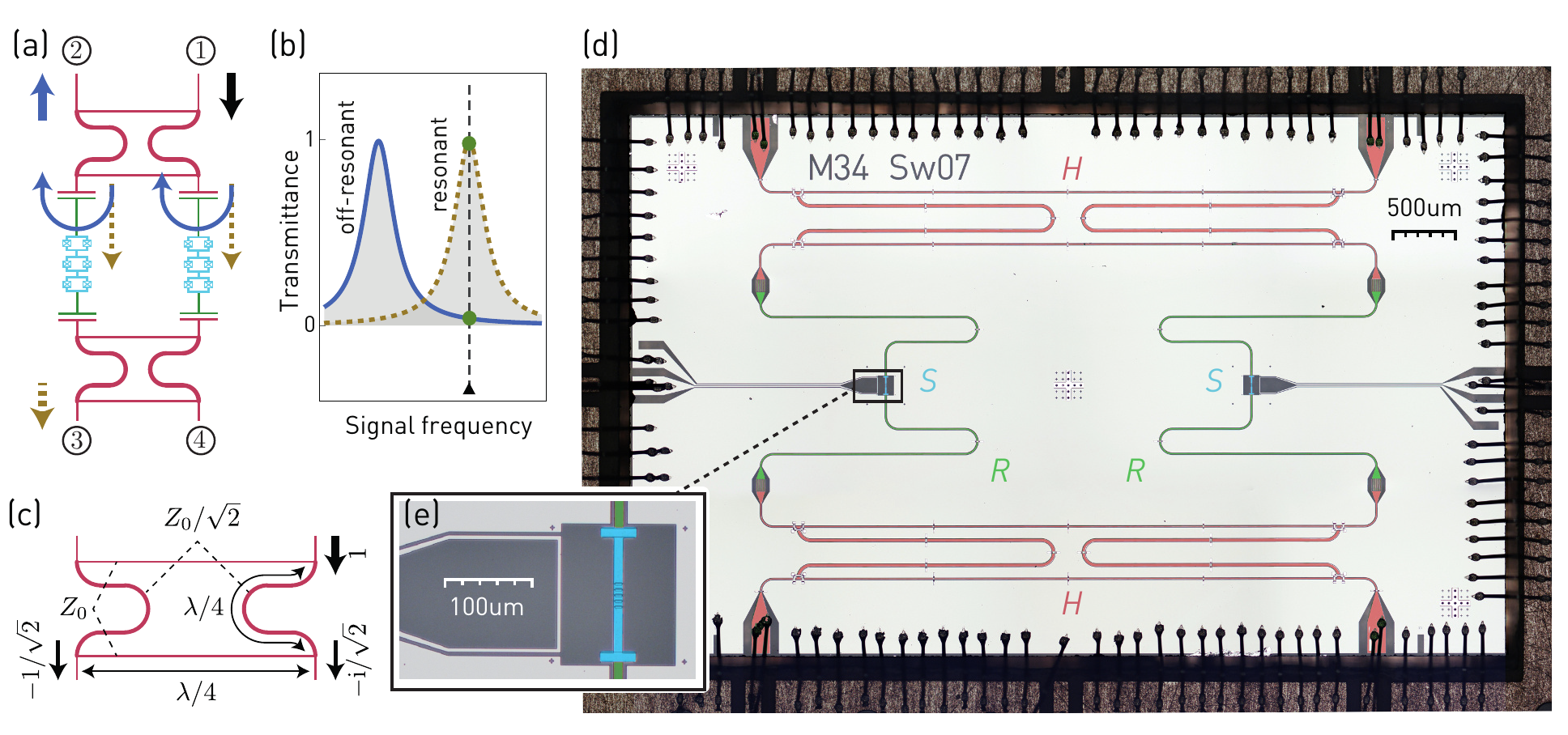}
\end{center}
\caption{(a) Schematic diagram of the switch illustrating the two switch states. The path taken by the signal is shown for the resonators off-resonant (solid arrows) and resonant (dashed arrows) with the signal. Input and output ports are labeled by $1,2,3,4$. (b) The theoretically expected transmittance of the resonators for the off-resonant and resonant switch states shown in (a). The signal frequency is indicated by the thin dashed line, the corresponding values of the transmittance by the green points. (c) Schematic diagram of the $\pi/2$ hybrid and the relevant $S$-parameters at the operating frequency. (d) False color micrograph of the sample mounted on its printed circuit board with the $\pi/2$ hybrids ({\it{H}}) indicated in red, the transmission line resonators ({\it{R}}) in green and the SQUID arrays ({\it{S}}) in blue. (e) Enlarged view of the SQUID array with the flux line loop visible on the left.}
\label{fig_switch_diagram}
\end{figure*}

The switch consists of two $\pi/2$ hybrid couplers \cite{Lang2013},\cite[pg. 343]{Pozar2011} connected in a Mach-Zehnder-like configuration with two independently tunable coplanar waveguide resonators \cite{Goppl2008,Palacios-Laloy2008} in the two arms, as shown in Fig.~\ref{fig_switch_diagram}(a). The resonance frequencies of the resonators are tuned by changing the inductance of  two arrays of $N=5$ SQUID loops placed in their centers (see supplementary material). The inductance of these arrays depends on a magnetic flux applied using two superconducting coils. The hybrid coupler is a structure consisting of four transmission line segments of equal length with pairwise equal characteristic impedances arranged in a square (see Fig.~\ref{fig_switch_diagram}(c)). Its effect on an incident microwave signal is analogous to that of an optical beam-splitter but the phases of the two resulting signals are phase-shifted with respect to each other by $\pi/2$. The hybrid lends itself to a straightforward implementation in the planar architecture of superconducting circuits where it has been used for quantum optics experiments with single microwave photons \cite{Lang2013}.

For the following discussion, we number the ports of the device according to Fig.~\ref{fig_switch_diagram}(a). The signal entering port 1 of the top hybrid is split equally into its two output ports and arrives at the resonators. When both of them are far detuned from the signal frequency, they act as nearly perfect reflectors (see Fig.~\ref{fig_switch_diagram}(b)). Due to the phase relation between the outputs of the hybrid, the two reflected parts of the signal recombine in port 2. This situation, corresponding to what we will call the \emph{off-resonant state} of the switch, is schematically illustrated by solid arrows in Fig.~\ref{fig_switch_diagram}(a). Conversely, in the \emph{resonant state}, both resonators are fully transmittive (see Fig.~\ref{fig_switch_diagram}(b)) and the signal arriving at the bottom hybrid will recombine in port 3, as shown by the dashed arrows in Fig.~\ref{fig_switch_diagram}(a).

An array of $N$ SQUID loops can be approximated as an inductor of inductance $L = N\hbar^2/(4e^2 E_J)$, where $E_J$ is the Josephson energy of each SQUID, related to the Josephson energies $E_{J1}$, $E_{J2}$ of the junctions comprising the SQUID and the magnetic flux $\Phi$ threading the loop by $E_J = \sqrt{E_{J1}^2+E_{J2}^2+2E_{J1}E_{J2}\cos(2e\Phi/\hbar)}$. The approximation by an inductor is valid only in the limit where the currents flowing across the junctions are small in comparison with their critical currents, that is, for sufficiently low signal powers. At higher powers, non-linear effects cause compression of the transmitted signal. Since the higher-order corrections to the SQUID array energy scale as $1/N^2$, using an array rather than a single SQUID loop improves the linearity of the device \cite{Palacios-Laloy2008,Eichler2014}. Our choice of $N=5$ (see Fig.~\ref{fig_switch_diagram}(e)) is a compromise between suppression of non-linear effects, which favors high $N$, and uniformity of the magnetic flux through the individual SQUID loops of the array, which is harder to achieve when the array is made longer.

\section{Measurement setup and device parameters}

The switch is fabricated on a $7\times 4\,\mathrm{mm}$ sapphire chip with the coplanar waveguide structures patterned by photolithography and etching of a niobium film (see Fig.~\ref{fig_switch_diagram}(d)). Air-bridges, made in an additional optical lithography step by deposition of aluminium and titanium, are distributed along the transmission lines to suppress potential parasitic modes on the chip. The SQUID arrays (see Fig.~\ref{fig_switch_diagram}(e)) are fabricated by electron-beam lithography and shadow evaporation of aluminium. A test chip with resonators in which the SQUIDs are replaced by an equally-sized niobium strip directly connecting the two resonator halves is measured in a dipstick at $4.2\,\mathrm{K}$ to determine the effective length of the resonator. The obtained resonance frequency of the fundamental mode is $\omega_{r0}/2\pi = 8.30\,\mathrm{GHz}$ and its 3dB bandwidth $\kappa_0/2\pi = 305\,\mathrm{MHz}$. In another test measurement at $4.2\,\mathrm{K}$, the operating frequency of the hybrid couplers is found to be $\omega_{h}/2\pi = 7.2\,\mathrm{GHz}$. The SQUID arrays consist of $N=5$ loops with a designed asymmetry ratio $E_{J1}:E_{J2} = 2:3$, chosen to yield a tuning ratio between the minimum and maximum $E_J$ of $1:5$. The resulting frequency tuning range of approximately $5.5-7.5\,\mathrm{GHz}$ is sufficient to fine-tune the device to its optimal operating point but not much larger than necessary. This helps to keep the sensitivity of the resonator frequencies to external magnetic flux noise low. The Josephson energies of the two junctions, extracted from a measurement of the flux-dependence of the resonator frequency (see supplementary material), are $E_{J1}/h = 1.06\,\mathrm{THz}$ and $E_{J2}/h = 1.54\,\mathrm{THz}$, respectively.

\begin{figure}
\begin{center}
\includegraphics[width=8.0cm]{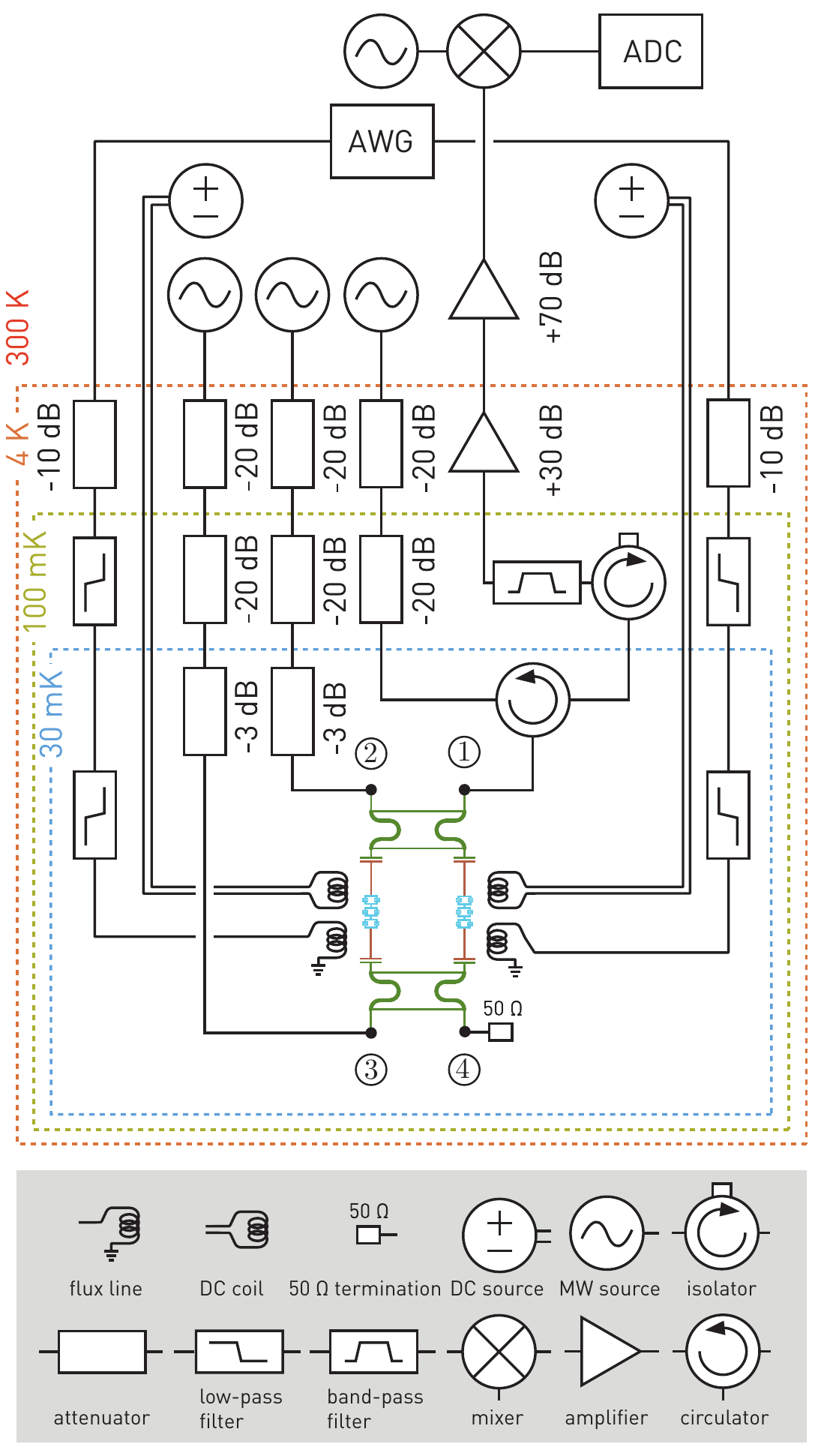}
\end{center}
\caption{Schematic diagram of the experimental setup with the relevant temperature stages of the dilution refrigerator indicated by the dashed boxes.}
\label{fig_setup_diagram}
\end{figure}

The experimental setup used in the initial characterization  measurements of the switch is shown in Fig.~\ref{fig_setup_diagram}. The device is mounted at the base plate of a dilution refrigerator and cooled down to a temperature of $30\,\mathrm{mK}$. Its ports 1, 2 and 3 are connected to three input lines and one output line through a circulator as indicated in the figure. The remaining port 4 is terminated with a $50\,\mathrm{\Omega}$ load. We measure the amplitude of the output signal using heterodyne detection with an intermediate frequency of $25\,\mathrm{MHz}$. To obtain the normalized $S$-parameters $S_{11}$, $S_{12}$ and $S_{13}$, the measured signal amplitude is calibrated using reference measurements at full transmission and reflection. These are taken at the base temperature of the cryostat, with the sample removed and with the end of microwave line~1 left open or with line~2~or~3 connected via a short coaxial through to line~1.

Magnetic flux bias of the SQUID arrays to tune the resonator frequencies is realized using two DC superconducting coils mounted on the sample holder underneath the two SQUID arrays. To achieve fast control of the flux, we use two on-chip flux lines (see Figs.~\ref{fig_switch_diagram}(d,e) and Fig.~\ref{fig_setup_diagram}) connected to an arbitrary waveform generator (AWG) with a bandwidth of $500\,\mathrm{MHz}$.

\section{Results}

\subsection{Characterization in the classical regime}

\begin{figure}
\begin{center}
\includegraphics[width=8.0cm]{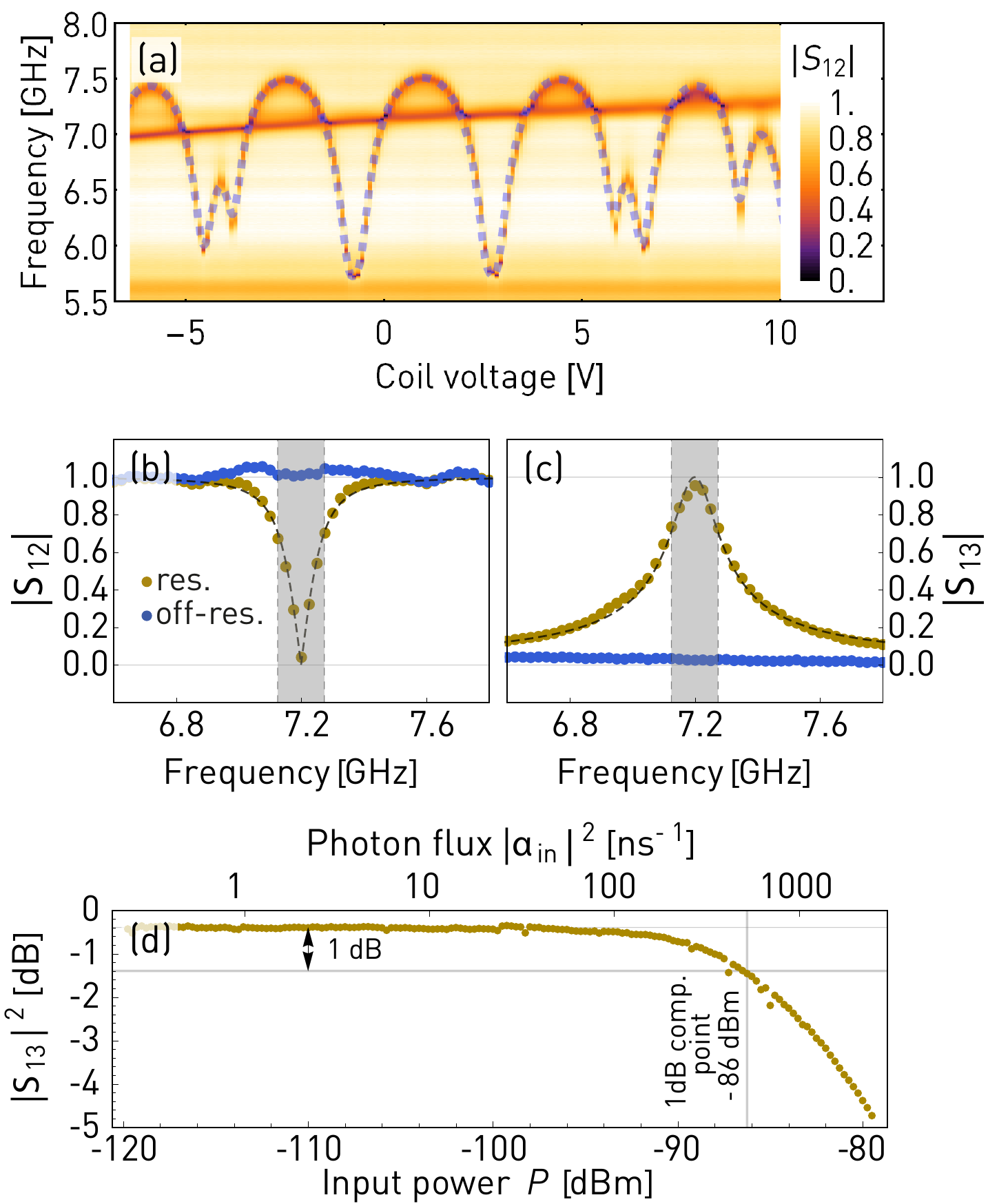}
\end{center}
\caption{(a) Transmission coefficient $S_{12}$ as a function of frequency and voltage applied to one of the flux bias coils. The dashed line represents a fit to a theoretical model (see supplementary material). Measured frequency-dependence of (b) the $S_{12}$ and (c) the $S_{13}$ transmission coeficients in the off-resonant state (blue) and the resonant state (yellow) of the switch. The values slightly larger than one result from imperfections of the calibration procedure. The gray band indicates the bandwidth of the device. (d) The $|S_{13}|^2$ transmittance on resonance (at $7.2\,\mathrm{GHz}$) in the resonant state of the switch as a function of the power (or, equivalently, the photon flux $|\alpha_{\mathrm{in}}|^2$) estimated at the input of the device.}
\label{fig_switch_spectrum}
\end{figure}

To characterize the device, we first measure the transmission coefficient $S_{12}$ as a function of frequency. We expect this coefficient to be close to unity when both resonators are far detuned from the signal and to decrease when either of the resonators is driven resonantly. By doing this measurement for different bias voltages applied to one of the two coils, we observe how the resonance frequencies change with magnetic field. The two lines visible in the measured dependence plotted in Fig.~\ref{fig_switch_spectrum}(a) correspond to the two resonators which couple to the coil with different strengths. The strongly coupled resonator shows approximately periodic behaviour of its frequency as expected from the dependence of the SQUID loop Josephson energy on magnetic flux. The imperfect periodicity is consistent with a small inhomogeneity of the magnetic field threading the different loops in the array. A fit based on a model taking this inhomogeneity into account (see supplementary material), shown in Fig.~\ref{fig_switch_spectrum}(a) by the dashed line, matches the data with good accuracy.

%
%

We set the signal frequency to $7.2\,\mathrm{GHz}$ -- the working frequency of the hybrid couplers -- and find the operating points of the switch by sweeping the bias voltages of both coils. We identify the two pairs of voltages for which the branching ratio $S_{12}/S_{13}$ is maximal and minimal. These operating points correspond to the off-resonant and the resonant state of the switch, respectively.

The ability of the switch to block signal transmission into the unwanted port can be characterized by the on/off ratios $|S_{12}^{(\mathrm{on})}/S_{12}^{(\mathrm{off})}|^2$ and $|S_{13}^{(\mathrm{off})}/S_{13}^{(\mathrm{on})}|^2$ between the transmitted powers in the resonant (superscript 'on') and off-resonant state (superscript 'off'). We evaluate these at the operating frequency of $7.2\,\mathrm{GHz}$ and obtain the values $28\,\mathrm{dB}$ and $32\,\mathrm{dB}$, respectively. We expect that this supression of approximately three orders of magnitude in power could be improved by increasing the tuning range of the resonators and optimizing the $\pi/2$-hybrid parameters.

We estimate the bandwidth of the switch by measuring the transmission coefficients $S_{12}$ and $S_{13}$ as a function of frequency at the two operating points of the device, obtaining the data shown in Fig.~\ref{fig_switch_spectrum}(b,c). We see that in the off-resonant state (blue), the coefficients are relatively insensitive to frequency changes. This is to be expected because the signal is detuned from the resonance frequencies of the two resonators by more than $1\,\mathrm{GHz}$, that is, by nearly ten times their linewidth. On the other hand, in the resonant state, the bandwidth of the device is determined by the linewidth of the tunable resonators. As seen in Figs.~\ref{fig_switch_spectrum}(b,c), the data closely match the fitted Lorentzian frequency response $|S_{13}|^2 = 1-|S_{12}|^2 = 1/(1+4(f-f_0)^2/\Delta f^2)$ shown by the dashed line. The center frequency $f_0 = 7.20\,\mathrm{GHz}$ and the bandwidth $\Delta f = \kappa/2\pi = 149\,\mathrm{MHz}$ are extracted from the fit.

To characterize the impedance-matching of the device, we measure the signal reflected from port 1 for both states of the switch. To account for reflections due to setup imperfections rather than the device itself, we compare the results with a reference measurement obtained when the device is removed and the connector of port 1 is terminated with a cryogenic $50\,\mathrm{\Omega}$ load, resulting in full absorption of the signal at the port. The power reflectance observed in this reference measurement is below $-10\,\mathrm{dB}$ within the $150\,\mathrm{MHz}$ wide band of the switch and we observe no systematic increase in this value when the device is connected, confirming that the reflections from the switch are small in comparison with the setup reflections.

We determine the linear operating range of the device by tuning it into the resonant state, probing it resonantly and observing the transmittance between ports $3$ and $1$ as a function of the applied power. The measured data presented in Fig.~\ref{fig_switch_spectrum}(d) show constant $|S_{13}|^2$ for powers lower than approximately $-100\,\mathrm{dBm}$ and a progressively higher compression as the power is increased. This non-linear effect can be quantified in terms of the 1 dB compression point -- the power at the input of the device at which the transmittance is reduced by 1 dB with respect to the low-power limit. To estimate the power level at the input port of the switch from the power generated by the microwave source, we measure the transmittivity of the input cables with a vector network analyzer. However, since this reference measurement is taken at room temperature and the actual transmittance is likely to be higher due to reduced losses of the microwave components at low temperatures, the power calculated in this way most likely underestimates the actual power by a few dB. As shown in Fig.~\ref{fig_switch_spectrum}(d), the 1 dB compression point extracted from the data is approximately $-86\,\mathrm{dBm}$, that is $2.3\,\mathrm{pW}$ or approximately $5\times 10^5$ photons per microsecond at $7.2\,\mathrm{GHz}$. The non-linearity of the device can also be estimated theoretically from the properties of the Josephson junctions forming the SQUID array (see supplementary material). 
A calculation using the experimentally estimated junction parameters yields the Kerr nonlinearity $K/2\pi = 28\,\mathrm{kHz}$ and the compression point $P_{\mathrm{cp}}=-81\,\mathrm{dBm}$ for this device. This agrees reasonably well with the measured value and we suspect the observed difference of $5\,\mathrm{dB}$ to be caused by the reduced losses in the microwave lines at low temperatures.

To study the switching speed at shorter time scales than those achievable with the magnetic field coils, we use on-chip flux lines (see Figs.~\ref{fig_switch_diagram}(d,e) and Fig.~\ref{fig_setup_diagram}) connected to the outputs of an arbitrary waveform generator capable of synthesizing waveforms with a~bandwidth of $500\,\mathrm{MHz}$. We first set the bias voltages of the coils to bring the switch to the resonant state in absence of any current through the flux lines. We then apply square pulses of varying amplitude to the flux lines to determine the voltages needed to switch the device to the off-resonant state. These are again found as the voltages corresponding to the maximum of the branching ratio $S_{12}/S_{13}$.

\begin{figure}
\begin{center}
\includegraphics[width=8.5cm]{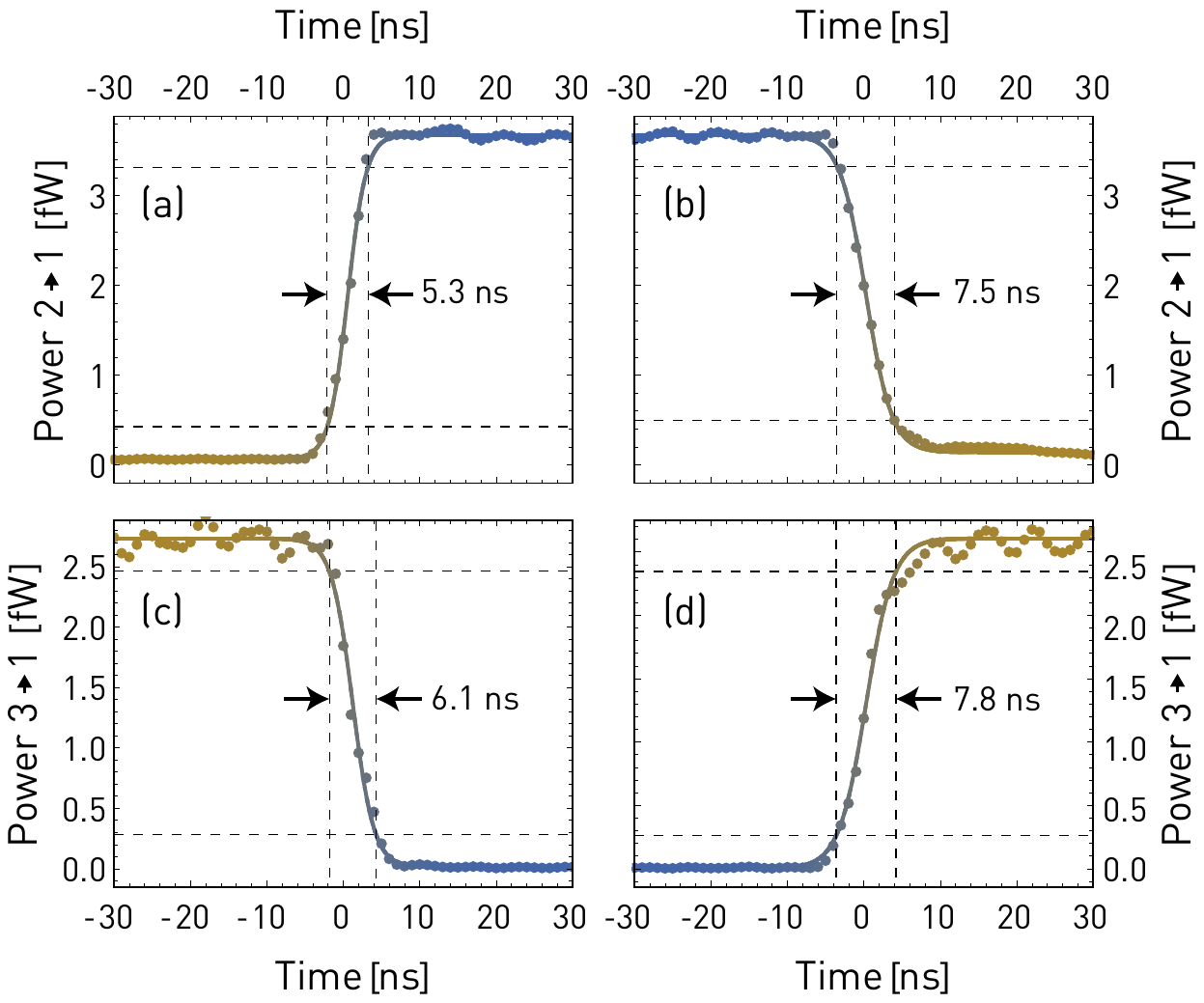}
\end{center}
\caption{Recorded waveforms of the (a) rising and (b) falling edges of the signal transmitted between ports 2 and 1 and between ports 3 and 1 (c,d). The colors of the points represent the two switch states as shown in Figs.~\ref{fig_switch_diagram} and \ref{fig_switch_spectrum}. Solid lines are fits to a $\tanh$ step used to estimate the rise/fall times. Dashed lines represent the 10\% and 90\% signal level used to extract the indicated switching times.}
\label{fig_plots_fluxLineSwitching}
\end{figure}

We set up a measurement where a continuous signal is applied to either one of the ports 2 and 3 and the output measured at port 1. Then we modulate the flux line voltages with a square pulse and record the downconverted waveforms (see Fig.~\ref{fig_plots_fluxLineSwitching}). For this measurement, we use a higher-bandwidth detection system with an intermediate frequency of $250\,\mathrm{MHz}$ and analog bandwidth of $500\,\mathrm{MHz}$ (see supplementary material). 

As shown in Fig.~\ref{fig_plots_fluxLineSwitching}, we fit a $\tanh$ step to the measured waveforms to extract the 10\%-90\% rise and fall times of approximately $5\,\mathrm{ns}$ and $7\,\mathrm{ns}$ for the signal applied at port $2$ and $8\,\mathrm{ns}$ and $6\,\mathrm{ns}$ for the signal applied at port $3$.

\subsection{Switching of quantum signals}

\begin{figure*}
\begin{center}
\includegraphics[width=16.5cm]{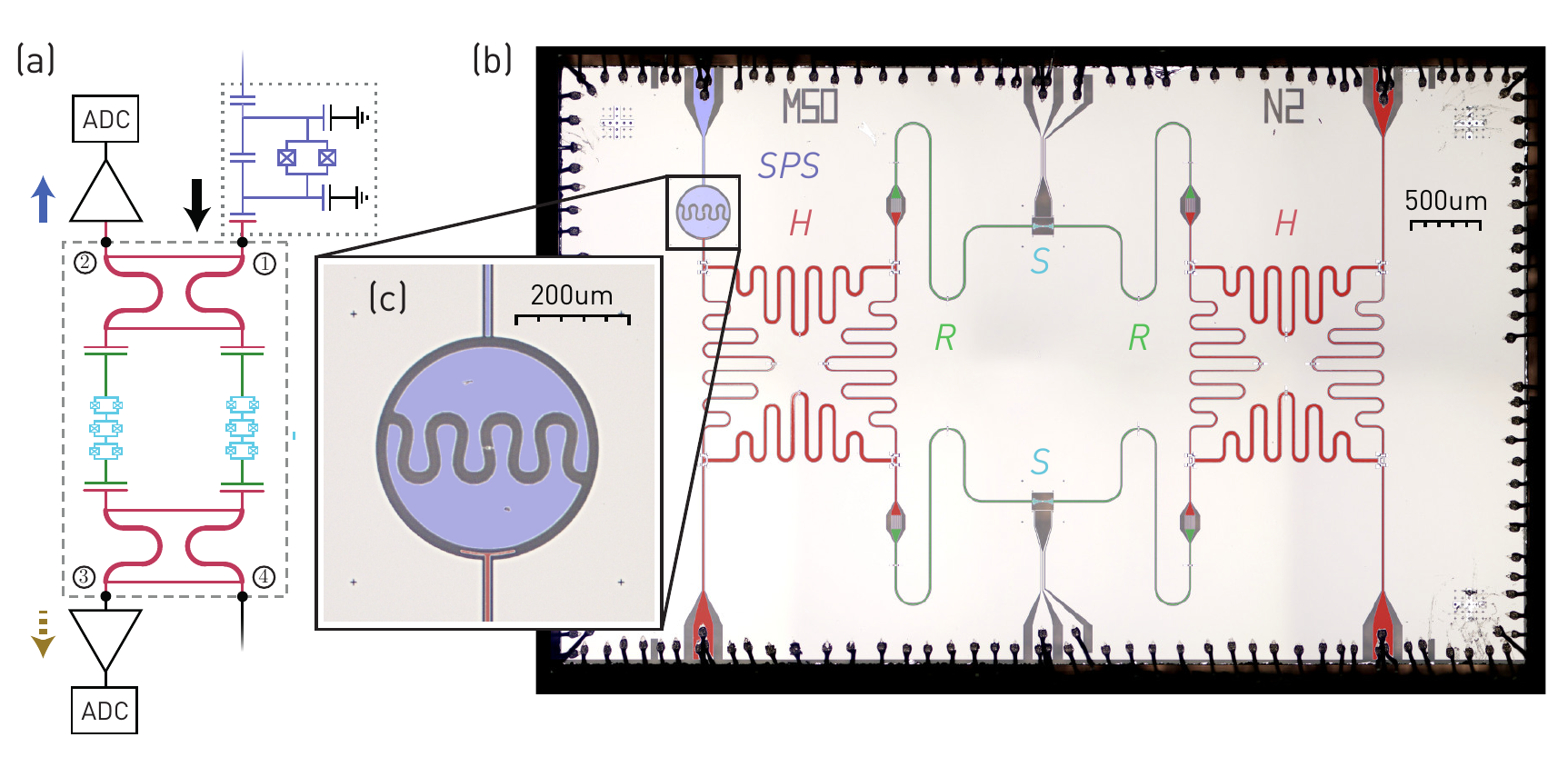}
\end{center}
\caption{(a) Diagram of the sample combining the switch (dashed box) with a transmon-based single photon source (dotted box). The detection chain represented by the triangular amplifier symbol uses a Josephson parametric amplifier for near-quantum-limited measurements. (b) False-color optical image of the sample wire-bonded on its printed circuit board, with the resonators ({\it{R}}) in green, $\pi/2$-hybrids ({\it{H}}) in red, SQUID arrays ({\it{S}}) in cyan and the single-photon source ({\it{SPS}}) in blue. The design of the $\pi/2$ hybrid is more compact than in the first generation switch to enable better scaling in future devices. (c) Enlarged view of the single-photon source with the two half-circular capacitor pads joined by the SQUID loop in the center. The weakly coupled drive line is shown in blue at the top and the strongly coupled output line in red at the bottom.}\vspace{0.5cm}
\label{fig_switchSPSphoto}
\end{figure*}

In order to simultaneously demonstrate the operation of the switch at the quantum level and its integrability with other quantum devices, we have fabricated a sample combining it with a single photon source on the same chip, as seen in Fig.~\ref{fig_switchSPSphoto}. The source is of the type demonstrated in Ref.~\cite{Peng2015}. It is implemented as a transmon-type superconducting qubit directly capacitively coupled to an input line of the switch and, with a weaker coupling, to a control line used to excite the qubit.

\begin{figure}
\begin{center}
\includegraphics[width=7.5cm]{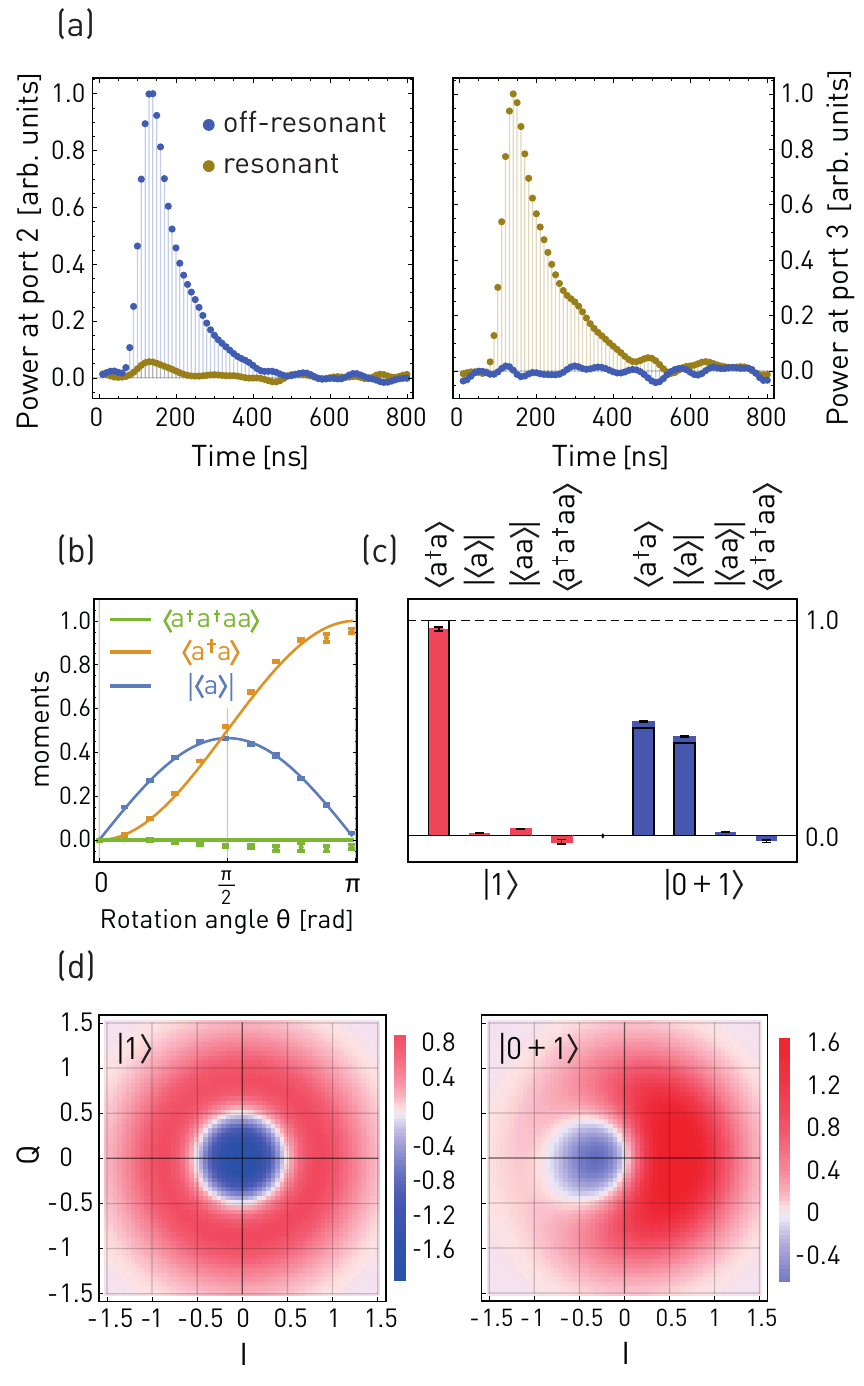}
\end{center}
\caption{(a) The single photon power waveforms measured at the output ports 2 (left) and 3 (right) in the resonant (yellow points) and off-resonant (blue points) state of the switch. (b) Calibration measurements of the moments $|\langle a\rangle|$, $a^\dagger a$ and $a^\dagger a^\dagger a a$ as a function of the rotation angle in the qubit initialization step. (c) Measured moments of a single photon state $|1\rangle$ (red) and a superposition state $(|0\rangle+|1\rangle)/\sqrt{2}$ (blue) transmitted through the switch. (d) Wigner functions corresponding to the moments from (c), obtained from them using a maximum likelihood method.}
\label{fig_switchSPSresults}
\end{figure}

The qubit is first excited by applying a short (approximately $10\,\mathrm{ns}$) microwave pulse, rotating its Bloch vector by an angle $\theta$ proportional to the pulse amplitude. This prepares the qubit in a superposition state of the ground state and the first excited state. When the qubit subsequently relaxes by photon emission into the input line of the switch, its state is mapped onto the corresponding superposition of the vacuum state $|0\rangle$ and the single-photon Fock state $|1\rangle$ of the propagating field. The photon state generated in this way has an exponential envelope with a time constant of approximately $90\,\mathrm{ns}$. The couplings to the qubit drive and switch input lines are designed to be sufficiently asymmetric such that the probability of the photon being emitted into the qubit drive line is only approximately 2\%. When designing the single photon source, we also engineered the direct coupling between the input and the output to be sufficiently weak to make the mean number of photons transmitted during the excitation pulse negligible. For this particular design, we estimate this number to be approximately $0.03$.

After routing the single photon state through the switch, we measure the signal power at its two outputs in both of its states. This is done by averaging the square of the measured voltage and subtracting the noise power obtained in an identical measurement with the signal source turned off \cite{Bozyigit2011}. The resulting waveforms are shown in Fig.~\ref{fig_switchSPSresults}(a) and demonstrate that the photon can be successfully routed to either one of the outputs. 

By calculating the moments of the measured single-shot voltages and systematically subtracting the reference values measured in the absence of the signal, as outlined for example in Ref.~\cite{Eichler2012}, we extract the moments of the photon mode $a$. Fig.~\ref{fig_switchSPSresults}(b) shows the moments $|\langle a\rangle|$, $\langle a^\dagger a\rangle$ and $\langle a^\dagger a^\dagger a a\rangle$ of the emitted field as a function of the rotation angle $\theta$ used in the qubit initialization step. We use these data to find the normalization coefficient for the moment measurements (see supplementary material). The measured moments are in close agreement with theory where the photon number $\langle a^\dagger a\rangle$ is expected to vary as $\sin^2(\theta/2)$ and the coherence $|\langle a\rangle|$ as $|\sin\theta|$.  

In particular, we analyze the Fock state $|1\rangle$ and the superposition state $(|0\rangle+|1\rangle)/\sqrt{2}$ (which we denote for the sake of brevity by $|0+1\rangle$) whose relevant moments are shown in Fig.~\ref{fig_switchSPSresults}(c). The first and second order moments agree well with the expected values $\langle a^{\dagger}a\rangle=1$, $|\langle a\rangle|=0$ for $|1\rangle$ and $\langle a^{\dagger}a\rangle=1/2$, $|\langle a\rangle|=1/2$ for $|0+1\rangle$. The slightly reduced value of $|\langle a\rangle|\approx 0.46$ for $|0+1\rangle$ can be explained as a result of dephasing of the photon-source qubit which leads to a prediction of $|\langle a\rangle|\approx 0.43$. As expected for a single-photon field, as opposed to a coherent or thermal state, the higher order moments such as $\langle a a\rangle$ or $\langle a^{\dagger}a^{\dagger} aa\rangle$ are close to zero.

Of particular interest is the normalized zero-time-delay intensity correlation function $g^{(2)} = \langle a^{\dagger}a^{\dagger}aa\rangle/\langle a^{\dagger}a\rangle^2$. Its value of $-0.03\pm 0.01$ for $|1\rangle$ and $-0.09\pm 0.02$ for $|0+1\rangle$ is very close to zero, showing nearly ideal anti-bunching of the switched microwave field. The fact that the value extracted from the experiments is slightly negative is most likely an artifact of the data analysis procedure which we ascribe to a non-vanishing thermal field in the off-measurement used as a reference. The non-classical nature of the switched signal is corroborated by the negative values of the Wigner function shown in Fig.~\ref{fig_switchSPSresults}(d), which is extracted from the measured moments by means of a maximum likelihood method \cite{Eichler2012}.

\vspace{0.5cm}
\section{Conclusions}

We have described the design, realization and characterization of an on-chip superconducting microwave switch and demonstrated that this device can be integrated in experiments at cryogenic temperatures, such as for example in the field of quantum information processing with superconducting qubits. The switching is based on interference effects in a circuit with two hybrid couplers and two resonators tunable by an externally applied magnetic field. The device studied here operates optimally at $7.2\,\mathrm{GHz}$, has a bandwidth of $150\,\mathrm{MHz}$, an isolation of $30\,\mathrm{dB}$ and is well impedance matched to $50\,\mathrm{\Omega}$. The resonators are designed to have low anharmonicity, allowing the switch to route approximately $5\times 10^5$ microwave photons per microsecond at its 1 dB compression point. An on-chip flux line allows us to perform switching on a nanosecond time-scale. We have integrated the device on a chip with a single-photon source and demonstrated its operation in the quantum regime by switching propagating Fock states and their coherent superpositions.

We expect this device to have a variety of uses in applications requiring fast multiplexing/demultiplexing of microwave signals directly at the mixing chamber stage of a dilution refrigerator. For example, it may be used for controlling and reading out multiple quantum systems with a small number of RF input and output lines, as an alternative to frequency multiplexing \cite{Chen2012f}. It should also present a convenient calibration tool for normalizing $S$-parameter measurements in the cryostat \cite{Ranzani2013} or characterizing linear detection chains in absolute photon-number terms by switching between the device-under-test and a suitable reference source \cite{Houck2007,Peng2015}. It can also be useful for building configurable networks for quantum optics experiments with microwaves \cite{Chen2011a}, remote entanglement distribution or quantum communication \cite{Munro2010,Collins2007}.

\section{Acknowledgements}

We would like to thank Patrice Bertet, Sebastian Probst, Benjamin Chapman and Eric Rosenthal for useful comments on the manuscript. This work is supported by the European Research Council (ERC) through the ``Superconducting Quantum Networks'' (SuperQuNet) project, by National Centre of Competence in Research ``Quantum Science and Technology'' (NCCR QSIT), a research instrument of the Swiss National Science Foundation (SNSF), and by ETH Zurich.

\end{document}